\documentclass{article}
\PassOptionsToPackage{numbers, compress}{natbib}
\usepackage[final]{neurips_2019}
\usepackage{amssymb,amsfonts,amsmath}
\usepackage{floatrow}
\usepackage{authblk}
\usepackage{caption}
\usepackage[utf8]{inputenc} 
\usepackage[T1]{fontenc}    
\usepackage{hyperref}       
\usepackage{url}            
\usepackage{booktabs}       
\usepackage{amsfonts}       
\usepackage{nicefrac}       
\usepackage{microtype}      
\usepackage[dvips]{graphicx}

\title{Spatiotemporal Information Processing with \\ a Reservoir Decision-making Network}

\author{
  Yuanyuan Mi\\
  Center for Neurointelligence, Chongqing University, Chongqing 400044, China,\\
  \texttt{miyuanyuan0102@163.com} \\

  Xiaohan Lin~~~~~Xiaolong Zou~~~~~Zilong Ji~~~~~Tiejun Huang~~~~~Si Wu\\
  School of Electronics Engineering and Computer Scienece, IDG/McGovern Institute for Brain Research, Peking University, Beijing 100871, China,\\
  \texttt{siwu@pku.edu.cn} \\
}

\newfloatcommand{figurebox}{figure}[\nocapbeside][\dimexpr(\textwidth-\columnsep)/2\relax]
\newfloatcommand{tablebox}{table}[\nocapbeside][\dimexpr(\textwidth-\columnsep)/2\relax]

\begin{document}

\maketitle

\begin{abstract}
Spatiotemporal information processing is fundamental to brain functions.
The present study investigates a canonic neural network model for spatiotemporal pattern recognition.
Specifically, the model consists of two modules, a reservoir subnetwork and a decision-making subnetwork.
The former projects complex spatiotemporal patterns into spatially separated neural representations,
and the latter reads out these neural representations via integrating
information over time; the two modules are combined together via supervised-learning using known examples.
We elucidate the working mechanism of the model and demonstrate its feasibility for
discriminating complex spatiotemporal patterns. Our model
reproduces the phenomenon of recognizing looming patterns in the neural system,
and can learn to discriminate gait with very few training examples.
We hope this study gives us insight into understanding how
spatiotemporal information is processed in the brain
and helps us to develop brain-inspired application algorithms.
\end{abstract}

\section{Introduction}
Spatiotemporal pattern recognition is fundamental to brain functions like recognizing moving objects or
understanding others' body language. In real neural systems,
all visual signals (optical flow) coming to the retina are continuous in time, generating
spike trains that propagate to the cortex. The brain is actually tasked with the problem of spatiotemporal pattern
recognition, rather than static image classification commonly studied in AI.
The key of spatiotemporal pattern recognition is to extract the spatiotemporal structures of image sequences.
Compared to our knowledge of how the brain extracts the spatial structures of visual images,
as is well documented by the receptive fields of neurons, our understanding of
how the brain processes spatiotemporal information is very limited.

Recently, motivated for AI applications, e.g., for video analysis,
a number of artificial neural network methods have been proposed to address
spatiotemporal pattern recognition. For instances,
mimicking the neural dynamics of the ventral pathway,
several authors extended convolutional neural networks (CNNs) by including spatiotemporal kernels,
which are trained to extract the spatiotemporal features of image sequences
\cite{Baccouche2011,Ji2013,Tran2015,Feichtenhofer2017}.
Mimicking the parallel processing of two visual pathways, Simonyan et al. proposed a two-stream model,
in which one stream extracts the features of static images and
the other stream the features of motion fields; and the two streams
are fused together to accomplish the recognition~\cite{Simonyan2016}.
These models mainly aim for AI, and it is unclear how they reveal the real neural mechanisms.

In the present study, we propose a novel neural model for spatiotemporal information processing.
Our model is mainly inspired by two sets of experimental findings.
Firstly, we note that in addition to the ventral and dorsal pathways, there exists a subcortical pathway from retina to superior colliculus (SC), which can recognize looming patterns rapidly~\cite{Gelder2008,Wei2015,Shang2015,Huang2017,Shang2018,Huang2019}. In this shortcut, there is no explicit feature extraction as in the ventral pathway, rather it works as reservoir computing~\cite{Jaeger2001,Yildiz2012,Legenstein2004}, that is, the large retina network holds the memory trace of external inputs via abundant recurrent connections between neurons, such that the spatiotemporal structure of a motion pattern is mapped into a specific state of the network; consequently, a linear network in SC can
read out the motion pattern. Secondly, we note that a decision-making network is good at integrating information over time needed for recognizing spatiotemporal patterns. In the classical experiments for motion discrimination, a monkey was presented with frames of moving dots embedded in strong background noises~\cite{Shadlen2001}. To judge the moving direction, the monkey needs to aggregate motion cues over frames. The modelling studies revealed that this evidence accumulating process was carried out by a decision-making network in LIP, where a group of neurons, each representing one choice, compete with each other via mutual inhibition to determine the final choice~\cite{Shadlen2001,Xiaojing2006}. The experimental data
for looming pattern recognition has suggested that the similar decision-making process may occur in SC~\cite{Shang2018,Huang2019}.

Inspired by the experimental findings, we investigate a canonical neural network model for
spatiotemporal pattern recognition. Specifically,
the model consists of two modules: a reservoir subnetwork and a decision-making subnetwork, referred to as a reservoir decision-making network (RDMN) hereafter (see Fig.~1).
Our model is different from others (those in \cite{Baccouche2011,Ji2013,Tran2015,Simonyan2016,Feichtenhofer2017}),
in terms of that
it does not extract features explicitly, but rather it employs a reservoir module
to project spatiotemporal patterns into spatially separable neural representations;
and moreover, it exploits a decision-making module to read out
these neural activity patterns via integrating information over time.
Using synthetic and real data, we demonstrate that our model works  well.

\section{The Model}
The basic structure of RDMN is shown in Fig.~1. In the below, we introduce the computational properties of each module in detail.
\begin{figure}[htb]
\begin{center}
\includegraphics[width=10cm]{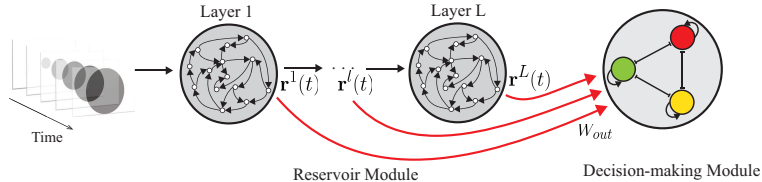}
\caption{The structure of RDMN. The network model consists of two modules,
a hierarchical reservoir subnetwork and a decision-making subnetwork. A spatiotemporal pattern
is first processed by the reservoir module, and then read out by the decision-making module.
The reservoir subnetwork consists of a number of forwardly connected layers, with each layer having a large number of recurrently connected neurons. The decision-making subnetwork consists of a number of competing neurons, with each
of them representing one class label. Each decision-making neuron receives
inputs from the reservoir module
and they compete with each other via mutual inhibition, with
the winner reporting the recognition result.}
\end{center}
\label{model-structure}
\end{figure}

\subsection{The decision-making module}
As shown in Fig.~1, the decision-making module consists of a number of competing neurons ($N_{dm}=3$ is shown), with each of them representing one class label. These neurons receive inputs from the reservoir module and compete with each other via mutual inhibition, with the winner reporting the discrimination result.
Denote $x_i$ the synaptic input received by neuron $i$, $r_i$ the corresponding neuronal activity, and
$s_i$ the synaptic current due to NMDA receptors.
The dynamics of neurons are written as (adopted and simplified from the mean-field model for decision-making in~\cite{Xiaojing2006}),
\begin{eqnarray}
x_i(t) & = & J_E s_i + \sum_{j\neq i}^{N_{dm}} J_M s_j + I_i,\\
r_i(t) & = & \frac{\beta}{\gamma} \ln \left[1+\exp\left(\frac{x_i-\theta}{\alpha}\right)\right], \\
\tau_s \frac{ds_i}{dt} & = & - s_i + \gamma(1-s_i) r_i. \label{slow-variable}
\end{eqnarray}
The synaptic input $x_i$ consists of three components: 1) $J_E s_i$, with $J_E>0$,
denotes the contribution of self-excitation (in a detailed network model, this corresponds
to the effect of excitatory interactions between neurons
representing the same category~\cite{Xiaojing2006});
2) $\sum_{j\neq i}^{N_{dm}} J_M s_j$ is the summed recurrent input from other decision-making neurons,
with $J_M <0$ indicating mutual inhibition between neurons; 3) $I_i$ is the feedforward
input from the reservoir module, whose form is determined through a learning process (see Sec.2.3).
The parameters $\beta, \gamma,$ and $\alpha$ control the shape of the nonlinear active function of neurons,
and $\theta$ the threshold. Eq.~(\ref{slow-variable}) describes the slow
dynamics of the synaptic current due to the activity-dependent NMDA receptors, which plays
a crucial role in decision-making~\cite{Xiaojing2006}. $\tau_s\gg 1$ is the time constant,
which controls the time window of integrating input information over time by a decision-making neuron.

\subsubsection{The mechanism}
Without loss of generality, we consider a simple case that
the decision-making subnetwork has two neurons, i.e., $N_{dm}=2$ for discriminating two spatiotemporal patterns.
The result can be straightforwardly extended to general cases of $N_{dm}>2$.
We study the network dynamics when both neurons receive the same constant
feedforward inputs, i.e., $I_i=I_0$, for $i=1,2$.
By varying the parameters, we find that the subnetwork can be at three types of
stationary state: 1) Low active state (LAS), in which both neurons are at the same low-level activity;
2) Decision-making state (DMS), in which one neuron is at high activity and the other at low activity, and
there are two such kind of states;
3) Explosively active state (EAS), in which both neurons are at the same high-level activity. Apparently,
DMS is suitable for decision-making, in which
only one neuron is active to report the discrimination result.

\begin{figure}[htb]
\begin{center}
\includegraphics[width=10cm]{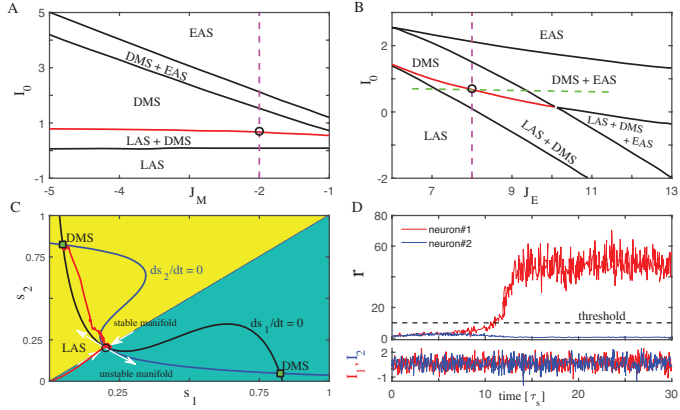}
\caption{The mechanism of decision-making.
(A-B) The phase diagram of the decision-making subnetwork with respect to:
(A) the feedforward input $I_0$ vs. the mutual inhibition $J_M$. $J_E=8$;
(B) the feedforward input $I_0$ vs. the self-excitation $J_E$. $J_M=-2$.
The stationary states of the subnetwork in different parameter regimes are shown.
The red lines denote the DM-boundary.
(C) The network dynamics when the parameters are on the DM-boundary (at the point denoted by
the black circles in (A-B)).
The red curve illustrates a typical example of the dynamics of decision-making neurons.
(D) An example trial of discriminating two temporal sequences.
Upper panel: the time courses of neuronal responses. Lower panel: two temporal sequences with slightly different means corrupted by large fluctuations, which are $I_1 = 0.7 + 0.6\xi_1(t)$ and $I_2 = 0.66 + 0.6\xi_2(t)$, with $\xi_1(t)$ and $\xi_2(t)$ denoting independent Gaussian white noises of zero mean and unit variance.
The black dashed line denotes the pre-defined threshold, and a decision is made once a neuron's activity crosses this threshold.
Other parameters are: $\alpha = 1.5$, $\theta=6$, $\beta=4$,$\gamma=0.1$,$\tau_s=100$.}
\end{center}
\label{working-mechanism}
\end{figure}

Fig.2A-B show the phase diagram of the subnetwork, which guide us to choose the network parameters.
For example, along the dashed vertical (pink) lines in Fig.2A-B (both $J_E$ and $J_M$ are fixed),
we observe that with the increase of the feedforward input $I_0$, the network dynamics
experience a number of bifurcations, that is, when $I_0$ is very small, the network
is only stable at LAS; as $I_0$ increases gradually,
the network becomes stable from at both LAS and DMS, to at only DMS, to at both DMS and EAS,
and to at only EAS. The parameters for decision-making
should be set at the regime where the network holds DMS as stable states.
Further inspecting the characteristics of neural dynamics suggests that
the optimal parameter regime should be at the bifurcation
boundary where LAS just loses its stability and DMS becomes the only stable state of the network
(indicated by the red lines in Fig.2A-B); hereafter, for convenience,
we call this boundary the decision-making boundary (DM-boundary).
On the DM-boundary, the network dynamics holds two appealing
properties: 1) since LAS is unstable, a feedfoward input with a little bias (e.g., $I_1>I_2$)
will drive the network to reach at one of DMS (e.g., neuron $1$ is active and neuron $2$ at low activity); 2)
because of supercritical pitchfork bifurcation,
the relaxation dynamics of the network is extremely slow, which
endows the network with the capacity of averaging out input fluctuations.

To elucidate the above properties clearly, let us investigate
the network dynamics by setting the parameters on the DM-boundary (at
the black circles in Fig.2A-B).
Fig.2C draws the nullclines of neuronal activities (the variables $s_i$, for $i=1,2$),
with their intersecting points corresponding to the unstable LAS and
two stable DMSs, respectively. According to the characteristic of supercritical pitchfork bifurcation,
the typical trajectory of the network state under the drive of a noisy input is as follows (illustrated by
the red curve in Fig.2C): starting from silence, the network
state is attracted first by the stable manifold of LAS; while approaching to LAS closely enough,
the unstable manifold of LAS starts to push the network state away, and this process is extremely slow
due to that the eigenvalue of the unstable manifold of LAS is close to zero at the supercritical pitchfork bifurcation point; once it is far away from LAS, the network state evolves rapidly to reach at one of DMSs.
Notably, due to slow evolving, the DMS the network eventually
reaches at is determined by integration of inputs over time, rather than by
instant fluctuations. This endows the network with
the capability of integrating temporal information over time in decision-making.

To confirm the above analysis, we perform a task of discriminating two temporal sequences
having slightly different means but corrupted with strong noises. To accomplish this task,
it is necessary to integrate inputs over time, so that large fluctuations are averaged out and
the subtle difference between means pops out.
Fig.2D presents a typical trial of the decision-making process: initially
the activities of two neurons are both low and intermingled with each other;
as time goes on, due to integration of inputs and competition via mutual inhibition,
the neuron receiving the larger input mean eventually wins.
To demonstrate that the optimal parameter regime is on the DM-boundary, we compare
network performances with varying parameter values, and observe
that when the parameters are away from the DM-boundary,
the discrimination accuracy degrades dramatically
(Fig.3A).

The above study reveals that the optimal parameter regime for decision-making should be on the DM-boundary.
Along this boundary, there are still flexibility to select the time constant $\tau_s$ and
the mutual inhibition strength $J_M$ (note that the DM-boundary is almost flat with respect to $J_M$, see Fig.2A).
We find that by varying $\tau_s$ or $J_M$ along the DM-boundary,
the network performance exhibits a speed-accuracy
trade-off (Fig.3B-C). This phenomenon is intuitively understandable.
With increasing $\tau_s$ and fixing other parameters, neurons
have a larger time window to average out temporal fluctuations, which
increases the discrimination accuracy but postpones
the decision-making time (measured by the activity of the wining neuron
crossing the threshold, Fig.2D).
Similarly, with increasing $J_M$, since the mutual inhibition between neurons becomes larger,
it tends to take a longer time for a neuron to win the competition (since the larger the mutual
inhibition, the stronger the suppressions between neuronal activities), which
postpones the decision-making time but improves the accuracy.
In practice, the values of $\tau_s$ and $J_M$ should match the statistics of input noises.

\begin{figure}[htb]
\begin{center}
\includegraphics[width=14cm]{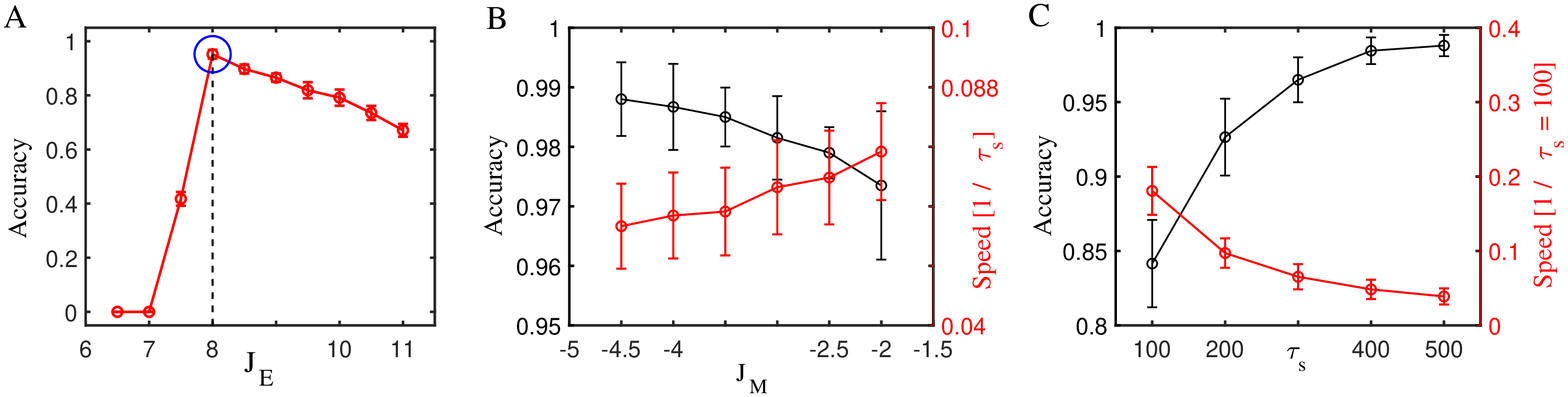}
\caption{Parameter setting for the decision-making module.
The network performances are evaluated by the task of
discriminating two temporal sequences given in Fig.2D.
The discrimination accuracy is measured by the rate that
the neuron receiving the larger input wins the competition.
The decision speed is measured by $1/t_{res}$, where
the decision-making time $t_{res}$ is measured as the moment when the activity of the wining neuron crosses
the pre-defined threshold (Fig.2D).
(A) The discrimination accuracy vs. the self-excitation strength $J_E$.
The value of $J_E$ varies along the dashed horizontal
green line in Fig.2B, with $J_E=8$ at the black circle on the DM-boundary.
(B-C) The speed-accuracy tradeoff of decision-making with respect to: (B) the mutual inhibition strength $J_M$
and (C) the time constant $\tau_s$. All results are obtained by averaging over $2000$ trials.
Other parameters are the same as in Fig.2.}
\end{center}
\label{network-performance1}
\end{figure}

The above analysis can be straightforwardly extended to the case when the
decision-making module has more than two neurons ($N_{dm}>2$) for
discriminating $N_{dm}>2$ spatiotemporal patterns.
We can compute the phase diagram of the decision-making subnetwork similarly
and obtain the optimal parameter regime accordingly.

\subsection{The reservoir module}
In the above, we have demonstrated that the decision-making module can
average out input fluctuations via its slow dynamics, and through competition,
discriminate temporal sequences of different mean values. However,
this module alone is unable to discriminate complex spatiotemporal patterns, e.g.,
it is unable to discriminate temporal sequences differentiable only in oscillation frequency
(since integrations of these sequences give the same mean value).
To be applicable to general cases,
it is necessary to add an additional module, the reservoir subnetwork (Fig.1).

Reservoir computing, also known as Liquid state machine or Echo state machine, has been proposed as
a canonical framework
for neural information processing~\cite{Jaeger2001,Yildiz2012,Legenstein2004}.
Its key idea is that rather than performing complicated computations,
a neural system projects external inputs of low dimensionality into activity patterns of a large size neural network, such that different input patterns become linearly separated as much as possible in the high dimensional space and hence
can be discriminated easily.
Here, as a pre-processing module before decision-making, we expect that
the reservoir subnetwork can map different spatiotemporal patterns into
spatially separated neural activities, so that
the decision-making module can discriminate them.

The reservoir subnetwork we consider consists of $L$ forwardly connected layers, and
neurons in the same layer are connected recurrently (Fig.~1).
Denote $x^l_i$ the synaptic current received by neuron $i$ in layer $l$, for $i=1,\ldots, N_l; l=1,\ldots, L$,
$N_l$ the number of neurons in layer $l$, and $r^l_i$ the neuronal activity.
The neural dynamics are given by,
\begin{eqnarray}
\tau_l \frac{dx^l_i}{dt} & = & -x^l_i +
\sum_{j\neq i}^{N_{l-1}}W^{l,l-1}_{i,j} r^{l-1}_j + \sum_{j=1}^{N_l}W^{l,l}_{i,j} r^l_j
+ \sum_{j =1}^{N_{in}} W_{i,j}^{l,0} I^{ext}_j \delta_{l,1}, \\
r^l_i & = & f(x^l_i),
\end{eqnarray}
where $\tau_l$ is the time constant of layer $l$, $W^{l,l-1}_{i,j}$ the forward connection from neuron $j$ at layer $l-1$ to
neuron $i$ at layer $l$, and $W^{l,l}_{i,j}$ the recurrent connection from neurons $j$ to $i$ at layer $l$.
$I^{ext}_j$ represents the external input, with $N_{in}$ the input dimensionality and $W_{i,j}^{l,0}$ the input connection weight; $\delta_{l,1}=1$, for $l=1$, and otherwise $0$, indicating that
only layer $1$ receives the external input. The nonlinear activation function of neurons is chosen to be
$f(x)=\tanh(x)$.
We set the parameters properly, so that each layer of the reservoir subnetwork holds
good dynamical properties, in terms of that: 1) starting from different initial states, the same external input will drive the network to reach the same stationary state, satisfying the so-called echo state property~\cite{Yildiz2012,Jaeger2001};
2) in response to different external inputs, the network states are significantly different,
realizing the so-called computing at the edge of chaos~\cite{Legenstein2004}.

Using simple examples, we illustrate the representation capacity of the reservoir module.
Fig.~4A presents the spectral analysis of neuronal responses in different layers
when a sequence of Gaussian white noises is applied as the external input.
It shows that along the layer hierarchy,
the dominating components of neuronal responses progress from high to low frequencies,
indicating that the frequency information of temporal inputs are separated across layers.
This implies if two temporal sequences have different frequencies, they can be
discriminated via neural activities across layers.
To confirm this property, we present two temporal sequences of different frequencies to the module
and observe that the neural activity patterns they generate are indeed separated (Fig.~4B).

\begin{figure}[htb]
\begin{center}
\includegraphics[width=10cm]{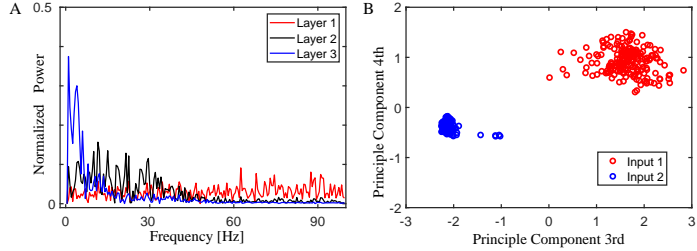}
\caption{Representation capacity of the reservoir module.
(A) Spectral analysis of neuronal responses across layers when a sequence of Gaussian white noises of zero mean and unit variance is applied as the external input. (B) Separation of neural representations in the reservoir module for
two temporal sequences of different frequencies, which are
$I^{ext,1}(t) = \sin(20\pi t)+\sin(200\pi t) + 0.1\xi_1(t)$ and
$I^{ext,2}(t) = \sin(40\pi t)+\sin(160\pi t) + 0.1\xi_2(t)$,
where $\xi_1(t)$ and $\xi_2(t)$ are Gaussian white noises of zero mean and unit variance.
Neural activities are projected onto their third and forth principle components, which have the largest contributions on
separating the two temporal sequences.
}
\end{center}
\end{figure}

\subsection{Combining two modules}
Using known examples, we can combine two modules together to
discriminate complex spatiotemporal patterns. The inputs
from the reservoir module to decision-making neurons are written as,
\begin{equation}
I_i=I^{*}_0+\sum_{l=1}^{L}\sum_{j=1}^{N^l} W^{dm,i}_{lj} r^{l}_j,
\end{equation}
where the summation runs over all neurons in the reservoir module, and
$W^{dm,i}_{lj}$ denotes the connection weight from neuron $j$ in layer $l$ of the reservoir subnetwork to neuron $i$ in
the decision-making module.
$I^{*}_0$ is the optimal feedforward input specified by the DM-boundary (see Fig.~2).
We find that by including this constant term (although it is not essential),
it helps to train the read-out matrix $\textbf{W}^{dm}=\{W^{dm,i}_{lj}\}$. In reality,
this may also have biological meanings, e.g.,
$I^{*}_0$ may represent a top-down signal controlling the operation of decision-making.

We optimize the read-out matrix  $\textbf{W}^{dm}$ using known examples.
According to the response characteristics of decision-making neurons as
observed in the experiment~\cite{Shadlen2001} and also in our model (Fig.2D),
we should set the target function for the activity of the winning neuron over time to be of the sigmoid-shape.
Since there is a monotonic relationship between the neuronal activity and its synaptic input (Eq.2), for
convenience, we directly
set the target function for the synaptic input to the wining neuron to be of the sigmoid-shape, which
guarantees to generate the sigmoid-shaped neuronal response.
Similarly, we set the target function for the synaptic input to a non-wining neuron
to be a constant. The read-out matrix is optimized through
minimizing the discrepancy between the target functions and the actual synaptic inputs received by decision-making
neurons, which is written as
$E=\frac{1}{2}\sum_{k=1}^{N_{dm}}\sum_{i=1}^{N_{dm}}\int_0^T dt \left[f^k_i(t)-x^k_i(t) \right]^2$,
where $f^k_i(t)$ denotes the target function for
decision-making neuron $i$ in response to spatiotemporal pattern $k$,
and $x^k_i(t)$ is the actual synaptic input received by the neuron obtained by evolving Eqs.1-3.
The number of decision-making neurons equals to that of spatiotemporal patterns, and
without loss of generality, we define that neuron $i$ encodes spatiotemporal pattern $i$.
For the wining neuron, the target function is $f^i_i(t)=J_E\left\{\tanh[b(t-T/2)]+1\right\}/2 + I^{*}_0$,
and for a non-winning neuron, the target function is $f^k_i(t)=J_M + I^{*}_0$, for $k\neq i$.
The input patterns last from $t=0$ to $T$.
With other parameters fixed by theory, we optimize the read-out matrix by
minimizing the error function $E$, and back propagation through time (BPTT) is used~\cite{Rumelhart1988}. We can also use
a biologically more plausible method, FORCE learning~\cite{Sussillo2009}, to get comparable performances.

Using the example of discriminating two temporal sequences (Fig.2D),
we can get an intuitive interpretation of what is learned by the read-out matrix.
Fig.4B shows that the two temporal sequences are well separated by
the third and forth principle components (PCs) of neuronal activities in the reservoir module.
Interestingly, we find that after learning, the read-out matrix also has larger overlaps with
these two PCs than with other components. This is reasonable,
as it gives rise to a larger difference between the
input means to decision-making neurons, which makes the discrimination task easier.
Thus, the read-out matrix learns to capture the key structure of neural representations
in the reservoir module necessary for discriminating the two temporal patterns.

\section{Model Performances}
\subsection{Looming pattern discrimination}
We first apply our model to discriminate looming patterns,
a type of spatiotemporal pattern that has been
widely used in the experiments to study the innate behavior of mice~\cite{Wei2015,Shang2015,Huang2017,Shang2018,Huang2019}.
A looming pattern is a sequence of light spot whose size increases continuously over time (Fig.5A),
and it mimics the approaching of a predator from the sky to a mouse.
The experiments have found that a mouse can discriminate looming patterns rapidly, and
triggers the innate response of pretending death if the size and speed of a looming pattern
are in the appropriate ranges.
Mimicking the experimental protocol, we construct three categories of looming patterns, which are: 1)
looming patterns having the appropriate speeds and sizes; 2) looming patterns having small or large speeds; 3) looming patterns having small sizes.
Fig.5B shows that our model performs this discrimination task very well,
and importantly, which is crucial for the innate response,
our model learns this task with very few trials.
\begin{figure}[htb]
\begin{center}
\includegraphics[width=10cm]{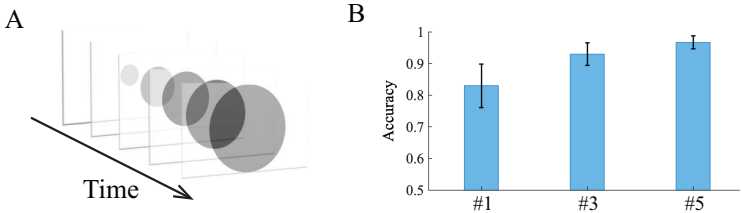}
\caption{
Looming pattern discrimination.
(A) An example of looming pattern.
(B) The accuracies of discriminating three categories of looming patterns.
$\#n$ means that $n$ trials for each class are used as training examples.
The results are obtained by averaging over $100$ testing trials.
}
\end{center}
\label{looming pattern}
\end{figure}
\subsection{Gait recognition}
We then apply our model to a real-world application, gait recognition, which is a typical example of
using spatiotemporal structures to discriminate objects.
To avoid that recognition relies on the side information of a subject (such as the height and shape of the subject), we extract skeletons from the images (see Fig.6).
The dataset we use consists of gaits of $17$ subjects, and each subject has $50$ trials,
with each trial lasting for $2$ seconds and containing $50$ image
frames~\footnote{$https://drive.google.com/drive/folders/1lbGD18Psb5jszMNHmzBc7bwl_hZkztG-?usp=sharing$}.
Three tasks of discriminating $5$, $10$, or $15$ people are performed, and
only one or five trials per person are used as training examples.
Our model is compared with LSTM, which has the similar idea of using neuronal recurrent interactions to hold the temporal information of inputs~\cite{Hochreiter1997}.
Fig.6 shows that the RDMN outperforms LSTMs in all tasks,
indicating that at least for few training examples,
the RDMN better captures the spatiotemporal structures of image sequences.
\begin{figure}[htb]
\begin{center}
\includegraphics[width=10cm]{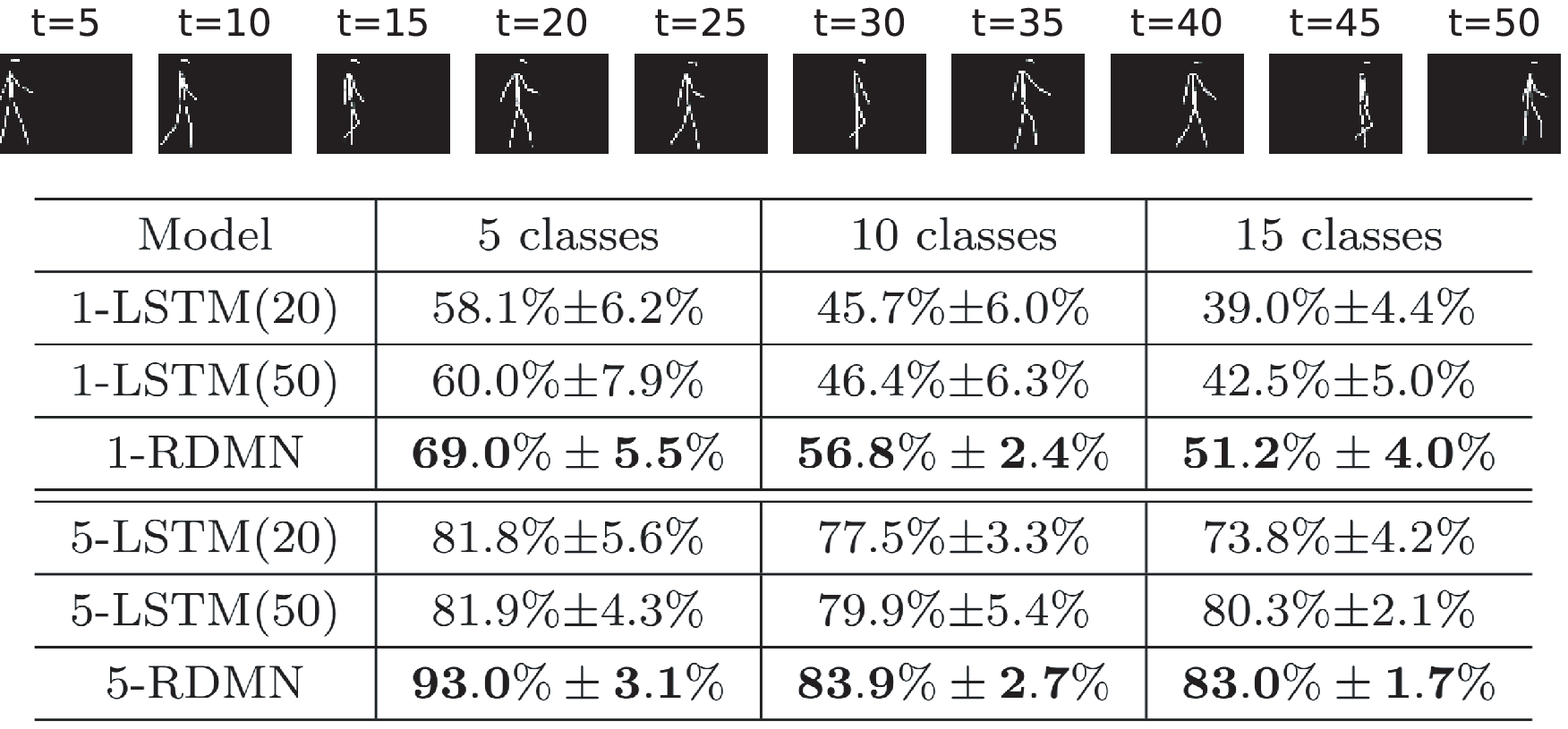}
\caption{
Gait recognition.
(A) An example of the gait of a subject.
(B) Comparing performances of different methods. Three tasks of discriminating $5, 10$ or $15$ subjects
are performed. 1-RDMN (1-LSTM) and 5-RDMN (5-LSTM) refer to that $1$ and $5$ trials per person are used as training examples, respectively.
LSTM(20) or LSTM(50) refer to that the number of units in the hidden layer is 20 or 50, respectively. In RDMN, the reservoir module has only one layer ($L=1$) with $N=1000$ neurons, and RDMN
has much less number of trainable parameters than LSTM.
}
\end{center}
\label{gait-recognition}
\end{figure}

\section{Conclusion}
In the present study, we propose a canonical model RDMN for spatiotemporal information processing
in neural systems. The model consists of two modules, a reservoir subnetwork and a decision-making subnetwork; the former aims to project complex spatiotemporal patterns into spatially separated neural representations,
and the latter aims to read out these neural activity patterns via integrating information over time.
We elucidate the working mechanism of the model and demonstrate its feasibility for
recognizing complex spatiotemporal patterns. Notably, although reservoir computing and decision-making,
as two canonical models for neural information processing,
have been widely studied in the literature, our work is the first one that links them
together as a unified model for spatiotemporal pattern recognition.
Furthermore, our work has other specific contributions, including elucidating in detail
the information integration mechanism of a decision-making module,
analyzing how temporal information is represented across layers in a reservoir module,
and developing a supervised-learning method to connect two modules.
Our model can be regarded as a simplified neural network that captures the fundamental structure of neural
systems for processing spatiotemporal information. In the case of the subcortical visual pathway,
the retina network can be regarded as the reservoir module (consider that visual signals mainly arrive at the fovea
and then spreads away to the periphery, the retina network can be approximated as a hierarchical reservoir network)
and SC the decision-making module.
Our model recognizes looming patterns successfully and can learn this task with very few trials.
We hope this study gives us insight into understanding how spatiotemporal information is processed in the brain.

Spatiotemporal pattern recognition is also a fundamental task in many AI applications.
Due to the lack of a good strategy for extracting and aggregating features from image sequences, it
remains an unresolved issue in the field. We apply our model to gait recognition and
find that our model outperforms LSTM and other approaches (see SI) under the setting of few training examples,
indicating that our model better captures the spatiotemporal structures of image sequences.
We will carry out comprehensive studies to evaluate our model on benchmark data in the future work.

\newpage

%
%
%
%


\small

\begin{thebibliography}{0}
\bibitem{Baccouche2011}
Baccouche M., F. Mamalet, C. Wolf, C. Garcia, and A. Baskurt,(2011) “Sequential deep learning for human action recognition,” Human Behavior Understanding, pp. 29–39.

\bibitem{Legenstein2004}
Bertschinger, N., Natschläger, T., \& Legenstein, Robert A. (2004). At the edge of chaos: real-time computations and self-organized criticality in recurrent neural networks. \emph{ Advances in neural information processing systems} 13–18. British Columbia, Canada: Vancouver.

\bibitem{Gelder2008}
De Gelder B, Tamietto M, Van Boxtel G J, et al. (2008) Intact navigation skills after bilateral loss of striate cortex[J]. \emph{Current Biology}, \textbf{18}(24): 1128-1129.

\bibitem{Feichtenhofer2017}
Feichtenhofer C., A. Pinz, and R. P. Wildes, (2017) “Spatiotemporal multiplier networks for video action recognition,” in \emph{CVPR}.

\bibitem{Hochreiter1997}
Hochreiter S, Schmidhuber J. (1997) Long Short-Term Memory. \emph{Neural Computation}, 9(8): 1735-1780

\bibitem{Huang2019}
Huang et al (2019) A Visual Circuit Related to Habenula Underlies the Antidepressive Effects of Light Therapy.
\emph{Neuron}, 102:1–15.

\bibitem{Huang2017}
Huang et al (2017) A retinoraphe projection regulates serotonergic activity and looming-evoked defensive behavior.
\emph{Nat. Comm.}, 8:14908.

\bibitem{Jaeger2001}
Jaeger, H. \ (2001) The ‘‘echo state’’ approach to analysing and training recurrent neural networks. \emph{ GMD report} 148, GMD — german national research institute for computer science.

\bibitem{Ji2013}
Ji S, Xu W, Yang M, et al. (2013) 3D Convolutional Neural Networks for Human Action Recognition. \emph{IEEE Transactions on Pattern Analysis and Machine Intelligence}, \textbf{35}(1): 221-231.

\bibitem{Rumelhart1988}
Rumelhart D E, Hinton G E, Williams R J. (1998)
Learning representations by back-propagating errors.
\emph{Cognitive modeling}, 5(3): 1.

\bibitem{Shadlen2001}
Shadlen M N, Newsome W T. (2001) Neural basis of a perceptual decision in the parietal cortex (area LIP) of the rhesus monkey. \emph{Journal of Neurophysiology}, \textbf{86}(4): 1916-1936.

\bibitem{Shang2018}
Shang C. et al (2018) Divergent midbrain circuits orchestrate escape and freezing responses to looming stimuli in mice.
\emph{Nat. Comm.}, 9:1232.

\bibitem {Shang2015}
Shang C. et al. (2015) A parvalbumin-positive excitatory visual pathway to trigger fear responses in mice.
\emph{Science}, 1472–1477.

\bibitem{Simonyan2016}
Simonyan K, Zisserman A. (2014) Two-Stream Convolutional Networks for Action Recognition in Videos. \emph{Advances in neural information processing systems}, 568-576.

\bibitem{Sussillo2009}
Sussillo D, Abbott L F. (2009) Generating Coherent Patterns of Activity from Chaotic Neural Networks. \emph{Neuron}, \textbf{63}(4): 544-557.

\bibitem{Tran2015}
Tran D, Bourdev L D, Fergus R, et al. (2015) Learning Spatiotemporal Features with 3D Convolutional Networks. \emph{international conference on computer vision}, 4489-4497.

\bibitem{Wang2013}
Wang H, Schmid C. Action Recognition with Improved Trajectories.(2013) \emph{international conference on computer vision},  3551-3558.

\bibitem{Wei2015}
Wei P, Liu N, Zhang Z, et al. (2015) Processing of visually evoked innate fear by a non-canonical thalamic pathway. \emph{Nature Communications}, 6756-6756.

\bibitem{Xiaojing2006}
Wong K F, Wang X J. (2006) A Recurrent Network Mechanism of Time Integration in Perceptual Decisions. \emph{The Journal of Neuroscience},  \textbf{26}(4): 1314-1328.

\bibitem{Yildiz2012}
Yildiz,I. B., Jaeger,H., Kiebel,S. J.(2012), Re-visiting the echo state property, \emph{Neural networks} \textbf{35}  1–9.

\end{thebibliography}
\end{document}